# Direct Wide-angle Measurement of Photonic Band-structure in a Three-dimensional Photonic Crystal using Infrared Fourier Imaging Spectroscopy

LIFENG CHEN,[1] MARTIN LOPEZ-GARCIA,[1] MIKE P. C. TAVERNE, [1] XU ZHENG, [1] YING-LUNG D. HO[1,*], JOHN RARITY[1,*]

[1]Department of Electrical and Electronic Engineering, University of Bristol, Merchant Venturers Building, Woodland
*Corresponding author: Daniel.Ho@bristol.ac.uk, John.Rarity@bristol.ac.uk

**We propose a method to directly visualize the photonic band-structure of micron size photonic crystals using wide angle spectroscopy. By extending Fourier Imaging Spectroscopy sensitivity into the infrared range we have obtained accurate measurements of the band-structures along the high-symmetry directions (X-W-K-L-U) of polymeric three-dimensional rod-connected diamond photonic crystals. Our implementation also allows us to record single-wavelength reflectance far field patterns showing a very good agreement with simulations of the same designs. This technique is suitable for the characterization of photonic structures working in the infrared and in particular, to obtain band-structure information of complete photonic band gap materials. © 2017 Optical Society of America**

Comprehensive measurements of the optical properties of photonic structures is fundamental to producing high quality micron size photonic devices. Among these devices, three-dimensional (3D) photonic crystal (PC) show promise for light guiding and trapping at sub-wavelength scales [1–3]. The most important optical properties of 3D-PC's are well defined by the so called photonic band structure (PBS) which describes the dispersion relation for propagating modes within the PC [3]. Therefore, by performing measurements of the PBS of a given photonic structure one can obtain a very accurate description of the optical performance of such devices [4]. One way to visualize the PBS is to obtain angle and polarization resolved reflectance/transmittance measurements of the 3D-PC at different wavelengths [5]. Bearing in mind that most PC are micron size structures one cannot rely on standard ellipsometry-based techniques for their study [6]. An alternative is Fourier Image Spectroscopy (FIS) [7] [8], also known as wide angle energy-momentum spectroscopy [9], which has been demonstrated to be an accurate method to investigate 3D-PCs [10] in the visible ranger but also to characterize the angular response of plasmonics [11] and bio-photonic [12] devices.

FIS relies on the spectroscopic examination of the image formed at the back focal plane of the objective lens which captures the scattering pattern of the device under study. Hence, imaging and measuring the PBS for devices working in the range of current CCD cameras (300 - 1000) is relatively straight forward [11]. That is not the case in the NIR (800 - 1700 nm) where 2D detector arrays present limited sensitivity and low resolution (hence low and fixed angular resolution for FIS measurements). Here, we present an alternative and more versatile implementation of FIS in the NIR (NIRFIS) that do not rely on 2D IR detector arrays. The technique described below allows the characterization of any far-field scattering pattern from a micron size device and could also be of interest in biological and medical applications [6] or surface analysis in the environmental sciences [13]. Moreover, angular resolution and sensitivity are disentangled in our implementation allowing high angular resolutions for a full description of the optical response of the 3D-PC. We demonstrate all these advantages by analyzing the wide angle optical response of Rod Connected Diamond (RCD) 3D-PC's, a microstructure known to allow full or partial photonic band gaps (PBG) in the NIR [14]. Measurement results are compared to simulations via finite-difference time-domain (FDTD) (based on Lumerical™ FDTD solutions® software) and plane wave expansion (PWE) (based on MIT Photonic Bands software [15]) methods.

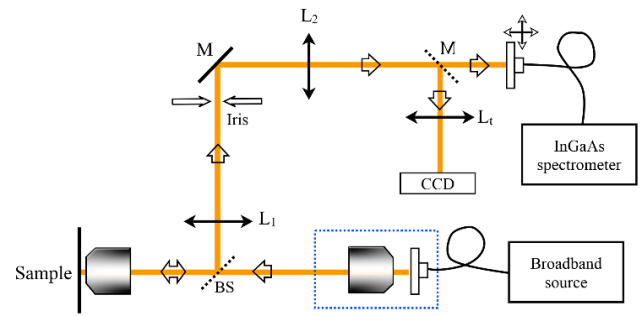

Fig. 1. Schematic diagram of the optical set-up based on Fourier Imaging Spectroscopy. Inside the blue dashed line frame is the collimator objective for the input beam. $L_1$ and $L_2$ are two confocal lenses of focal length $f_{L1} = f_{L2} = 300$ mm. $L_t$ is a tube lens for the camera of focal length $f_{Lt}= 150$mm. $L_1$ is located at its focal distance away from the back focal plane of the sample objective lens. A 50% reflection and transmission infrared beam splitter is mounted before $L_1$. The lens $L_2$ is positioned 300 mm from the detection plane, which is an x-y motorized translation stage attached with a multimode fiber.

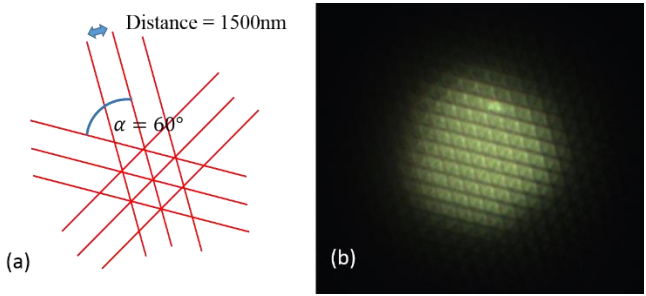

Fig. 2. (a) The design of polymeric hexagonal grating with pitch distance 1500 nm, rotation angle $\alpha = 60°$. (b) A microscope top view image of the hexagonal grating with pitch distance 1500nm.

Fig. 1 shows a sketch of our NIRFIS implementation. The output of a fiber coupled (Ø105 μm, 0.10 NA) broadband white light unpolarized source (Bentham Ltd. WLS100 300 nm – 2500 nm) is collimated by a low magnification objective lens (4× Olympus Plan Achromat Objective, 0.10 NA). The beam is then focused onto the sample by a high magnification objective lens (60× Olympus Plan Fluorite Objective) with NA = 0.9. The collection path consists of a 4f system formed by two achromatic lenses ($f_{L1} = f_{L2} = 300$ mm). The 4f system is placed at focal distance f from the back focal plane (BFP) of the objective lens. This ensure that the BFP image is projected into detection plane where it is spectrally analyzed by raster scanning with a multimode fiber (Ø200 μm, 0.22 NA) mounted on an x-y translation stage (Thorlabs motorized translation stage, resolution > 10 μm) and connected to a fiber coupled InGaAs based

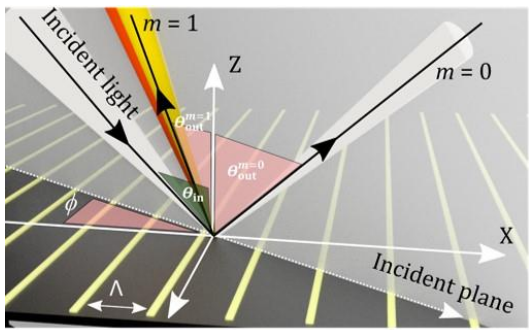

Fig. 3. Schematic drawing of a grating and incident plane, including angles of incidence, $\theta_{in}$, and reflection, $\theta_{out}$, the diffraction order ($m = 0$, ±1, ±2 …), the pitch distance of grating Λ, and rotation angle between the incident light plane and the grating groove $\phi$.

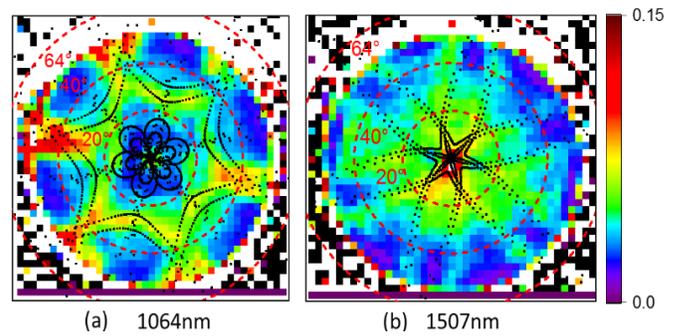

Fig. 4. Reflection back focal plane image from the hexagonal grating in Fig. 2 at wavelengths λ = 1064 nm (a) and 1507 nm (b). Red dashed lines indicate the collection angle at 64°, 40° and 20°. Black dashed lines are calculated diffraction boundary of the gratings for different wavelength.

spectrometer (NIRQuest512, Ocean optics Ltd., spectral range 900 - 1700 nm).

Prior to the characterization of any 3D-PhC it is necessary to perform a characterization of the system using a microstructure with a well-known optical response. Here we opt for a hexagonal polymeric grating structure fabricated on a glass substrate, which consists of three sets of parallel lines has rotated by 60° to each other as shown in Fig. 2. The reasons to use this type of sample are twofold. First, the device can be fabricated via direct laser writing (DLW), in our case using a commercial implementation (Photonic Professional, Nanoscribe, Germany). Secondly, the optical properties of this structure can be simple predicted from 1D gratings. The diffraction angles for the different orders can be calculated analytically using Bragg diffraction law for gratings as shown below and illustrated in Fig. 3:

$$\boldsymbol{k}_{in} = \boldsymbol{k}_{out} + m\boldsymbol{G}_x , \qquad (1)$$

where **m** is diffraction order (0, ±1, ±2 …), $\boldsymbol{G}_x$ is the reciprocal vector for the grating ($|\boldsymbol{G}_x| = 2\pi/\Lambda$ and Λ the grating period). The wave vector $\boldsymbol{k}_{in}$ of the incident light is a function of the incident angle $\theta_{in}$ and the azimuthal angle $\phi_{in}$ between the incident plane and the grating orientation $\hat{x}$ :

$$\boldsymbol{k}_{in} = |\boldsymbol{k}_0| \sin\theta_{in} \cos\phi_{in}\hat{x} + |\boldsymbol{k}_0| \sin\theta_{in} \sin\phi_{in}\hat{y} . \qquad (2)$$

Without any grating structure the reflection field is uniform from Fresnel surface reflection. With a structured surface the primary features we see in FIS are associated with *disappearance* of reflection when incident light is preferentially coupled into high angle diffraction orders and surface modes. Thus our observation of an edge in the pattern is associated with the appearance of a diffraction order at $\theta_{out} < 90°$ a small peak in reflectance can also be observed in high quality structures due to wood's anomaly [16]. As this is a subtraction effect the edge appears in the same plane as incident plane with $\phi_{in} = \phi_{out}$ . The equation governing the appearance of grating orders is obtained from equations (1) and (2) at the limit $\theta_{out} = 90°$.

$$\theta_{in} = \sin^{-1}\left(1 + \frac{m\lambda}{\Lambda \cdot \cos\phi}\right), \qquad (3)$$

where $\lambda$ the wavelength of incident light. On the fabricated samples we choose a pitch Λ = 1500 nm which will show various diffraction edges in the spectral region of interest (900-1700nm) within the angles available for the objective lens (NA = 0.9, $\theta_{max}$ = 64.1°).

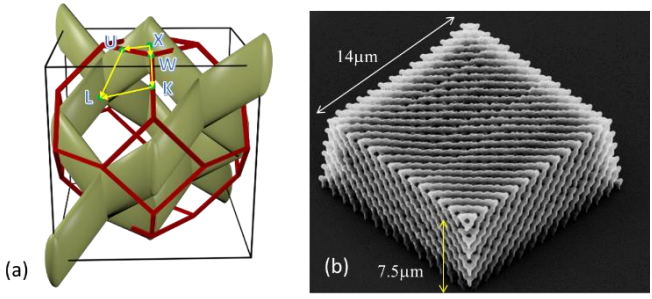

Fig. 5. (a) The unit cell of RCD structure, red line indicates its first Brillouin zone, and X, W, K, U and L are symmetry points in irreducible Brillouin zone. (b) A SEM image of a polymeric RCD structure, with size 14 μm in horizontal and 7.5 μm in vertical.

The measurement is processed by scanning the back focal plane of the objective lens at the detection plane with interval step size 0.15 mm between each collection point. There are 41 columns and 41 rows for the hexagonal grating measurement. The reflectance measured at each step is spectrally normalized against a reflectance standard, in this case a silver mirror. As a result, single wavelength snapshots for the angular diffraction pattern of the grating can be generated. Fig. 4 shows two examples at wavelengths 1064 nm and 1507 nm. Both patterns show reflection peaks, with reflectivity at around 10%. The reflection peaks lie along the diffraction boundary lines calculated from equation 3 (black dashed line) indicating a good performance of the setup. The red dashed lines indicate the collection angle at 64°, 40° and 20° on the image. A clear boundary between measurement data and the noise (white and black region) is the limit of the collection angle in this measurement system, and this limit corresponding to calculation is around 55°.

To fully demonstrate the potential of this technique, we investigate the angular response of Rod-connected diamond (RCD) [14] structures. The RCD structure is illustrated in Fig. 5, this structure is based on the diamond lattice replacing the bonds of adjacent atoms with rods (a), and its symmetry points in the Brillouin zone are X (0,0,1), U, L (1,1,1), K and W. This 3D-PhC structure is of technological interest because it shows the widest PBG among all photonic structure designs [17]. In this paper the polymeric structures were fabricated via the same DLW method used for the calibration gratings. The SEM picture of a fabricated polymer RCD structure is presented in Fig. 5b.

The measured reflection data (unpolarized) are presented in Fig. 6 as a color contour plot from 900 - 1700 nm wavelength plotted against X-W-K-L-U in symmetry directions. The reflectivity from low to high is correlated to the rainbow color from purple to red. The calculated PBS is plotted as colored lines overlapped on top, which predicts possible bandgaps at all symmetry directions. The bright green (40% reflectivity) patterns are associated with the high reflectivity at the stop band where light propagation is prohibited, whereas the dark blue and purple region are high transmittance where light can propagate through the structure. In particular, the fundamental bandgap appearing between the 2$^{nd}$ and 3$^{rd}$ bands shows up as a wide reflection peak. The angle dependent shift of center wavelength from 1700nm to 1300nm and the ~100 nm gap roughly agree with the simulation predictions. Hence it is clear that our technique is giving direct access to the bandstructure of a photonic crystal across a range of wavelengths and propagation directions.

We can also extract the Fourier plane image from the measured data at individual wavelengths. Fig. 7a-d show single wavelength snapshots of the measured far field pattern in reflectance for the fabricated structures at wavelengths, 1300, 1400, 1500 and 1600 nm. For comparison, Fig 7e-f shows FDTD calculation of the far field reflectance patterns expected for structures with the parameters described above. The symmetry points (X-W-K-L-U) (Fig. 5a) in the k-space of the RCD structure are labeled on the back focal plane images. The figures show how the reflection features evolve as wavelength changes from 1300 nm to 1600 nm in both measurement and simulation results.

At 1300 nm wavelength there are strong reflection peak features at the W symmetry point and a lower reflection feature at the U point in both measurement and simulation (Fig. 7a and Fig. 7e), and these features are more clear in the PBS plot (Fig. 6). At wavelength 1400nm, both U and W symmetries show similar reflectivity (40% in measurement, Fig. 7b), agreeing with simulation (Fig. 7f) within the 40° range. Fig. 6 shows a clear fundamental bandgap reflection peak (green area) around the 1500 - 1600nm region. The bandgap shifts and shrinks as the collection angle changes. These features are clearly seen in Fig. 7c where the reflection peak (40% reflectivity) area has expanded and covers from normal incidence (Γ-X direction) to W and U points and in Fig. 7d (1600 nm) where the fundamental reflection peaks have dropped (to 20%) and are now restricted to low angles and for high angles at the L point. Comparing measurement results to simulations we see reflection features showing good agreement within 40° range. Lower reflectivity features seen at higher angles (above 40°) are probably due to the finite structure size in the horizontal plane (in simulation the structure is infinitely wide). All measurements at high incident angles are limited by the objective lens in the sample plane, the coating of the objective is optimized for visible wavelength range thus numerical aperture (NA) shrinks (from 0.9 to 0.8) in our measurement wavelength range (900 - 1700 nm), and this results in about 3 degrees lost in the L direction (Fig. 6).

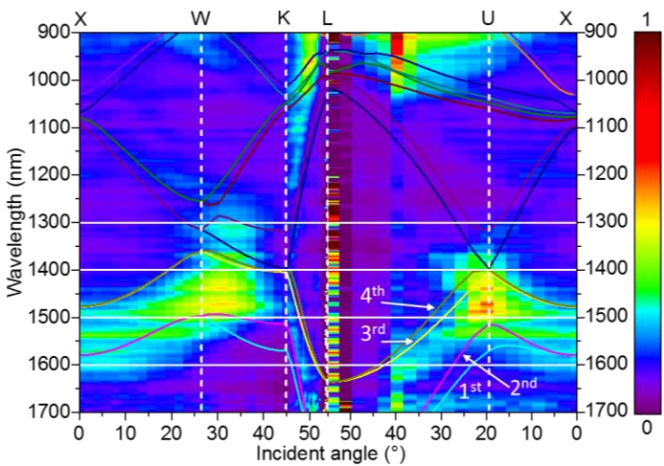

Fig. 6. The measured angular reflection spectrum mapped with photonic band diagram calculations (color lines) from 0° to 55°, symmetry points (shown in Fig. 5a) (X-W-K-L-U) are transferred and labelled (vertical white dashed lines) to incident angles (Γ-X as normal incident). Horizontal white line shown wavelength snapshots positions in Fig. 7. The first 4 bands are labeled and followed with arrows.

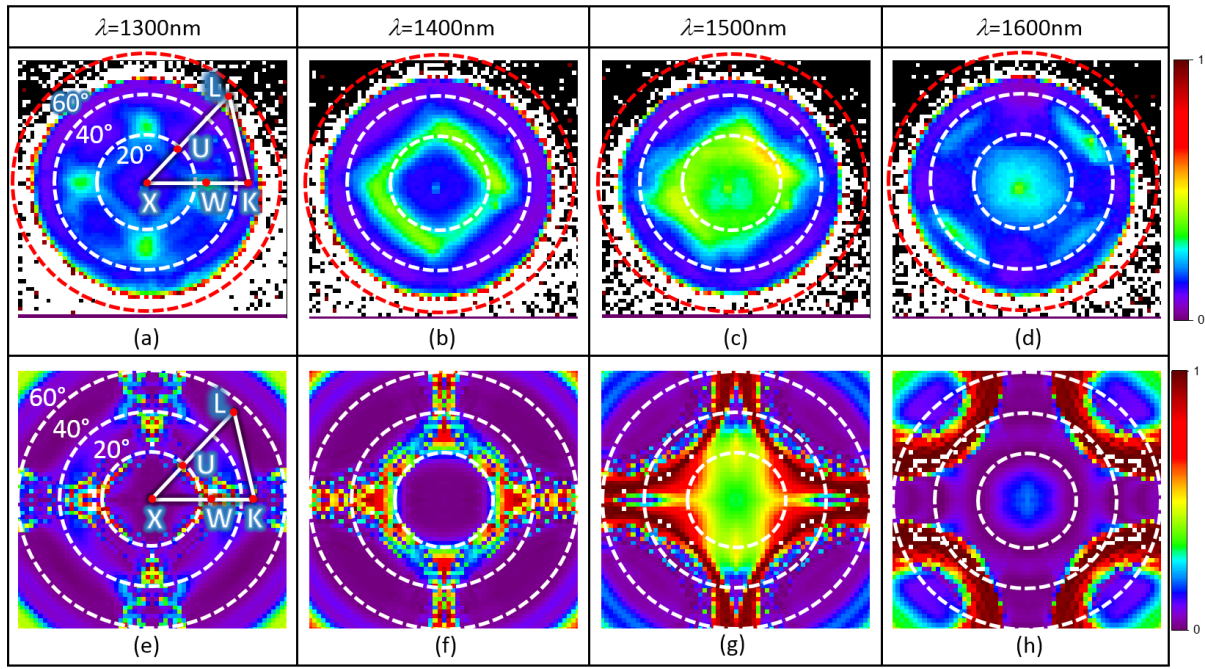

Fig. 7. Reflection back focal plane image from polymer RCD structure at wavelength (a) 1300 nm, (b) 1400 nm, (c) 1500 nm and (d) 1600 nm in measurements and its correspondent FDTD simulation reflection calculations (e), (f), (g) and (h). Dashed lines showing the angular contour lines on the image, 60°, 40° and 20°. Solid white line with red dots indicates the k-space symmetry points transferred to corresponding angles, these points are X, W, K, L and U.

In conclusion, we have designed and built a Fourier imaging spectroscopy (FIS) system using a fiber coupled spectrometer and accessing high angle scattering through a high NA objective lens. We use the system to visualize wavelength snapshots of reflectivity or transmittance in the Fourier plane and the band structure of photonic crystals. A hexagonal grating and an RCD structure, fabricated via DLW method, are used in the measurement tests and compared to simulations. The hexagonal grating measurement shows the observation angle of the setup extends up to 55°. In the results of RCD photonic crystal structure, a full photonic band structure is measured, and mapped to the irreducible Brillouin zone of the structure, showing good match to the trends of fundamental bands in the wavelength range 1400 – 1600 nm. Wavelength snapshots in the Fourier plane in 2D compare well with FDTD calculations, showing similar features. Hence we have shown the capability of visualizing and characterizing complex photonic crystals and their photonic bandstructure in the near infrared wavelength range. In future the resolution of the measurements and imaging could be improved further by modifying the magnification of imaging system and using smaller core fiber or smaller measurement intervals.

In other infrared applications, for instance in the medical field, by using mini-sized optical components, this setup has potentials to work portably as an individual instrument, and perform measurement on the go. Moreover, our setup with FT-IR could also be used for surface analysis of proteins or large crystals, which require angular dependent spectrum measurements in the infrared region.

**Funding.** ERC advanced grant 247462 QUOWSS; EPSRC grant EP/M009033/1.

**Acknowledgment.** This work was carried out using the computational facilities of the Advanced Computing Research Centre, University of Bristol, Bristol, U.K. JGR and Y-LDH acknowledge financial support from the ERC advanced grant and EPSRC grant.